\documentclass[aps,prb,twocolumn,showpacs,groupedaddress]{revtex4}
\usepackage{graphicx}
\begin{document}

\title{Self-trapped electrons and holes in PbBr$_{2}$ crystals}
\author{Masanobu Iwanaga}
\affiliation{Graduate School of Human and Environmental Studies, 
Kyoto University, Kyoto 606-8501, Japan}

\author{Junpei Azuma}
\author{Masanobu Shirai}
\author{Koichiro Tanaka}
\affiliation{Department of Physics, Graduate School of Science, 
Kyoto University, Kyoto 606-8502, Japan}

\author{Tetsusuke Hayashi}
\affiliation{Faculty of Integrated Human Studies, 
Kyoto University, Kyoto 606-8501, Japan}

\date{\today}

\begin{abstract}
We have directly observed self-trapped electrons and holes 
in PbBr$_{2}$ crystals with electron-spin-resonance (ESR) technique. 
The self-trapped states are induced below 8 K by two-photon interband 
excitation with pulsed 120-fs-width laser light at 3.10 eV. 
Spin-Hamiltonian analyses of the ESR signals have revealed 
that the self-trapping electron centers are the dimer molecules of 
Pb$_2$$^{3+}$ along the crystallographic $a$ axis and 
the self-trapping hole centers are those of Br$_2$$^-$ 
with two possible configurations in the unit cell of the crystal. 
Thermal stability of the self-trapped electrons and holes 
suggests that both of them are related to the blue-green luminescence band 
at 2.55 eV coming from recombination of spatially separated electron-hole 
pairs. 
\end{abstract}


\maketitle

\section{INTRODUCTION}
Lead halide, PbBr$_2$ and PbCl$_2$, crystals decompose photochemically 
under ultraviolet (UV) light irradiation at room temperature.~\cite{Verwey} 
At low temperatures, intense photoluminescence (PL) with large Stokes shift 
is induced instead of the photochemical 
decomposition.~\cite{Gruijter2,Liidja} 
Because ionic conductivity is suppressed at low temperatures, 
it is thought that local lattice deformation just after photoexcitation 
does not induce further successive desorption of 
halogen ions and aggregation of lead ions. 
The local lattice deformation is most likely related to the PL 
at low temperatures. 
These phenomena imply the strong electron-phonon interaction in 
PbBr$_2$ and PbCl$_2$. 

By studying the PL properties at low temperatures below 30 K in PbBr$_2$, 
it has been strongly suggested that bound and free electron-hole 
($e$-$h$) pairs intrinsically relax into 
spatially separated pairs of a self-trapped electron (STEL) 
and a self-trapped hole (STH).~\cite{Iwanaga1,Iwanaga2} 
The relaxation probably results from the strong interaction of both 
electrons and holes with acoustic phonons.~\cite{Sumi} 
To obtain further insight into the relaxation of $e$-$h$ pairs, 
it is significant to investigate structurally the 
lattice-relaxed localized states with electron-spin-resonance (ESR) 
technique in PbBr$_2$ photoirradiated below 30 K. 

Structures of localized electronic states have been investigated so far 
in other ionic crystals. 
In x-ray irradiated crystals of alkali halides, ESR experiments 
have revealed that the holes localize on two nearest-neighbor halogen ions 
and form the dimer molecules of (halogen$_2$)$^-$ 
(Refs.\ \onlinecite{Song} and \onlinecite{Silsbee}). 
The top of the valence band is composed of the $np$ states in $X^-$ ions 
($X=$ F, Cl, and Br for $n=2$, 3, and 4, respectively), 
and the formation of the dimer-molecular STH is mainly attributed to 
the covalent bond of the $np$ states.~\cite{Schoemaker2} 
On the other hand, holes in cubic PbF$_2$ irradiated with 
$\gamma$ ray or neutrons localize on Pb$^{2+}$ ions and 
form Pb$^{3+}$ centers below 77 K.~\cite{Pb3+2} 
Some calculations~\cite{Nizam,Fujita}  for cubic PbF$_2$ 
have indicated that the top of valence band is mainly 
composed of the $6s$ states in Pb$^{2+}$. 
The STH structure is closely related to the valence-band structure. 
Moreover, Pb$^{2+}$ ions can be effective hole traps and form Pb$^{3+}$ 
centers in Pb-doped alkali halide.~\cite{Pb3+1} 

STEL's were first observed in PbCl$_2$ irradiated with 
x ray~\cite{Nistor1} or $\gamma$ ray~\cite{Hirota} at about 80 K; 
the STEL's form the dimer molecules of Pb$_2$$^{3+}$ which are 
complementary to $X_2{}^{-}$ (V$_{\rm k}$ center) in alkali halides, 
and the configuration is due to the conduction band composed of the $6p$ 
states in Pb$^{2+}$ ions~\cite{Fujita} 
and to the covalent bond of the $6p$ states. 
It was proposed that STH's in PbCl$_2$ form V$_{\rm k}$-type Cl$_{2}$$^{-}$ 
centers,~\cite{Nistor2} but the existence has been disputable. 
Although no definite experimental evidence simultaneously observing 
both STEL's and STH's has been reported so far,
PbCl$_2$ and PbBr$_2$ are the candidates for yielding 
them.~\cite{Iwanaga3} 
The coexistence of STEL's and STH's gives the evidence that both electrons 
and holes strongly interact with acoustic phonons,~\cite{Sumi} and provides 
the experimental foundation for the further study on the uncommon relaxation 
including spontaneous $e$-$h$ separation.~\cite{Iwanaga1,Iwanaga2,Iwanaga3} 

In order to clarify the relaxed states of $e$-$h$ pairs in PbBr$_2$, 
we have measured ESR signals induced with pulsed 120-fs-width and 3.10-eV 
laser light below 8 K. Spin-Hamiltonian analyses of the ESR signals 
have revealed that both electrons and holes get self-trapped, and 
respectively form dimer molecules of Pb$_2$$^{3+}$ and of Br$_2$$^-$ 
as the self-trapping centers. 
We present the properties of ESR signals of PbBr$_2$ and the thermal 
stability in Sec.\ \ref{results}, analyze the structures of 
the electron- and hole-trapping centers with spin Hamiltonians 
in Sec.\ \ref{analysis}, and discuss the correlation of 
STEL's and STH's with luminescence in Sec.\ \ref{discussion}. 

\section{EXPERIMENTAL PROCEDURES}
Single crystal of PbBr$_2$ was grown with the Bridgman technique from 
99.999\% powder purified under vacuum distillation. 
The crystal of orthorhombic D$_{2h}^{16}$ (Refs.\ \onlinecite{Wyckoff} and 
\onlinecite{Wells}) was cut in the size of 3$\times $3$\times $3 mm${^3}$ 
along the right-angled crystallographic $a, b$, and $c$ axes. 

The crystal was sealed in a transparent quartz capillary with a transparent 
quartz rod and was fixed in a microwave cavity with a guide for light 
injection. The microwave cavity was a rectangular TE$_{103}$ cavity 
resonator with the quality factor $Q=3000$. 
The sample was photoirradiated below 8 K with the second harmonics 
(pulsed 120-fs-width, 1-kHz, and 3.10-eV light) generated from a 
regeneratively amplified Ti:sapphire laser; the average power of the incident 
light was about 100 mW/cm$^2$ on the sample surface and the photoirradiation 
time was typically one hour. The incident photons induce the two-photon 
interband transition, create $e$-$h$ pairs almost uniformly in the crystal, 
and produce measurable ESR signals within one minute. 
To avoid optical bleaching of the ESR signals, 
the sample after photoirradiation was kept dark during the measurement. 

The photoirradiated sample was measured below 8 K with ordinary ESR technique 
in X-band range; the resonant microwave frequency was 9.378$(\pm 0.008)$ GHz. 
Rotation-angle dependence of the ESR signals was measured by rotating the 
crystal around the $a$, $b$, and $c$ axes. 
Thermoluminescence (TL) under pulse annealing was measured through 
the transparent quartz rod attached to the sample.
The total TL during the pulse annealing were directly detected by a 
photomultiplier, and the TL spectra at various pulse-annealing 
temperatures were measured with a charge-coupled device camera 
equipped with a grating monochromator. 
Raising- and lowering-rates of temperature under pulse annealing were 
about 5 K$/$s, and the sample was typically annealed for one second 
at each annealing temperature. 
The ESR spectra after pulse annealing were measured below 8 K. 

\section{EXPERIMENTAL RESULTS}
\label{results}
\subsection{Properties of ESR spectra\label{ESRresult}}
\begin{figure}
\includegraphics[height=55.7mm,width=75mm]{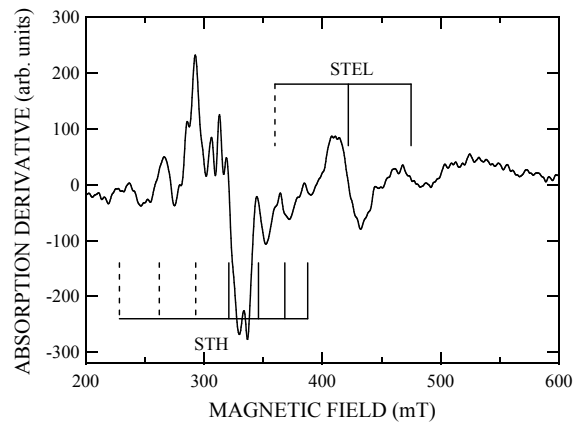}%
\caption{Typical ESR spectrum measured at 5 K after photoirradiation at 5 K. 
Magnetic field vector is in the $ac$ plane and the directon corresponds 
to 65$^{\circ}$ in Fig.\ \ref{STH}. Microwave frequency is 9.385 GHz. 
STEL and STH are described in Sec.\ \ref{ESRresult}. Solid lines indicate 
the prominent ESR positions. Dashed lines correspond to the ESR positions 
calculated from the spin Hamiltonian in Sec.\ \ref{analysis}.}
\label{ESR}
\end{figure}

Figure \ref{ESR} shows a typical ESR spectrum measured at 5 K after 
photoirradiation at 5 K. 
The solid lines in Fig. \ref{ESR} indicate the prominent ESR positions and 
the dashed lines do the ESR positions calculated from the spin Hamiltonians 
in Sec.\ \ref{analysis}. 
The signals are classified into three groups by the rotation-angle 
dependence and the thermal profiles: 
(i) The ESR signals named as STEL survive up to 130 K under pulse annealing. 
(ii) The signals named as STH disappear under pulse annealing above 30 K. 
(iii) Though the several signals in 280-340 mT cannot be well resolved 
because of the overlap with the STH signals, they survive above 100 K 
and almost disappear around 130 K under pulse annealing. 
The signals named as STEL and STH respectively correspond to self-trapped 
electrons and holes, as analyzed in Secs.\ \ref{AnaSTEL} and \ref{AnaSTH}. 

ESR signals around 330 mT under UV light irradiation at 77 K were 
reported by several authors.~\cite{Arends,Gruijter,Kerssen} 
However, the ESR signals named as STEL and STH have not been reported 
to our knowledge. Because of the overlap of the ESR signals, 
the spectrally well-resolved STEL and STH signals are mainly restricted to 
the region higher than 340 mT; in the region the intensity ratio of the STEL 
signals is $4:1$ and that of the STH signals is $3:2:1$. 
The STEL signals show linear response to the microwave power up to 10 mW 
even at 5 K. On the other hand, the STH signals saturate 
for the microwave power higher than 0.01 mW at 6 K. 
The reduced line width of the STEL signals, 
$\Delta B = (g/g_{0})\Delta B_{0}$ 
is estimated to be $14(\pm 1)$ mT where $g$ is the $g$ value of the STEL, 
$g_{0}$ the free-electron $g$ factor, and $\Delta B_{0}$ the line width 
estimated from the experimental data; the $\Delta B$ is about three-times 
broader than $\Delta B=5.5$ mT in PbCl$_2$ (Ref.\ \onlinecite{Nistor1}). 

\begin{figure}
\includegraphics[height=50.5mm,width=80mm]{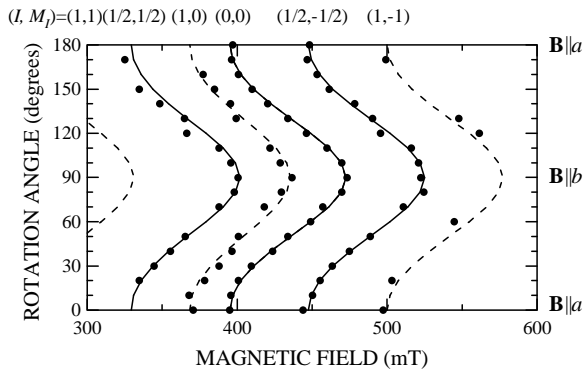}%
\caption{Photoinduced STEL signals (closed circle) vs rotation angle 
measured at 8 K. Microwave frequency is $9.371(\pm0.001)$ GHz. 
Magnetic field vector {\bf B} is in the $ab$ plane and the rotation axis 
is the $c$ axis. $(I, M_I)$: the total nuclear spin of the STEL 
centers and the magnetic quantum number.
Solid and dashed lines: fitting lines derived from the spin Hamiltonian 
described in Sec.\ \ref{AnaSTEL}.}
\label{STEL}
\end{figure}

Figure \ref{STEL} presents the rotation-angle dependence of the STEL signals 
(closed circle) at 8 K; magnetic field vector \textbf{B} is in the $ab$ 
plane and the rotation axis is the $c$ axis. 
Index $(I, M_I)$ denotes the pair of the total nuclear 
spin of the STEL center and the magnetic quantum number. 
The intensities of the ESR signals with indexes $(1, 0)$ and 
$(1, -1)$ are more than ten-times as weak as the $(0, 0)$. 
The intensity ratio between $(0, 0)$, $(1/2, 1/2)$, and $(1/2, -1/2)$ 
is about $4:0.9:1$. Solid and dashed lines fit the experimental data 
and are derived from the spin Hamiltonian (\ref{Ham}) 
in Sec.\ \ref{AnaSTEL}. A similar rotation-angle dependence of the STEL 
in PbCl$_2$ shows the twofold splits when the $\mathbf{B}$ is in the $bc$ 
and the $ac$ planes;~\cite{Nistor1,Hirota} 
the splits come from the two configurations reflecting the crystallographic 
symmetry. However, the explicit split 
has not been observed in PbBr$_2$ when the $\mathbf{B}$ is in the $bc$ and 
the $ac$ planes. Probably, the broad line width in PbBr$_2$ hides the split. 

\begin{figure}
\includegraphics[height=45.8mm,width=80mm]{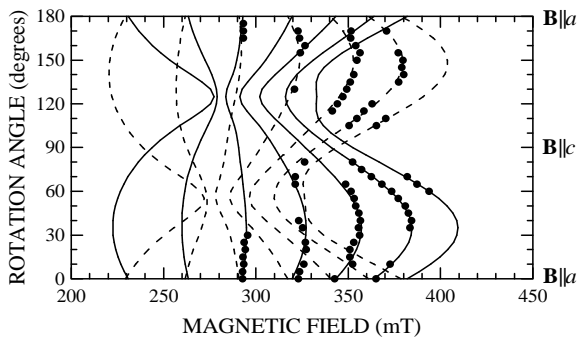}%
\caption{Photoinduced STH signals (closed circle) vs rotation angle 
measured at 8 K. Microwave frequency is $9.383(\pm0.002)$ GHz. 
Magnetic field vector {\bf B} is in the $ac$ plane and the 
rotation axis is the $b$ axis. Solid and dashed lines: 
fitting lines derived from the spin Hamiltonian described 
in Sec.\ \ref{AnaSTH}.}
\label{STH}
\end{figure}

Figure \ref{STH} displays the rotation-angle dependence of the STH signals 
(closed circle); the magnetic field vector $\textbf{B}$ is in the $ac$ plane 
and the rotation axis is the $b$ axis. 
The signals are picked up from the ESR spectra as shown in Fig.\ \ref{ESR}.
There exist the two series each of which has more than three resonances. 
The two series are mirror symmetric for the $bc$ plane with each other, 
reflecting the crystallographic mirror symmetry. 
Solid and dashed lines are derived from the spin Hamiltonian describing 
the hole center composed of two-equivalent Br$^{-}$ ions, and fit the STH 
signals. 

\subsection{Thermal stability of ESR signals and thermoluminescence}
\begin{figure}
\includegraphics[height=46.5mm,width=80mm]{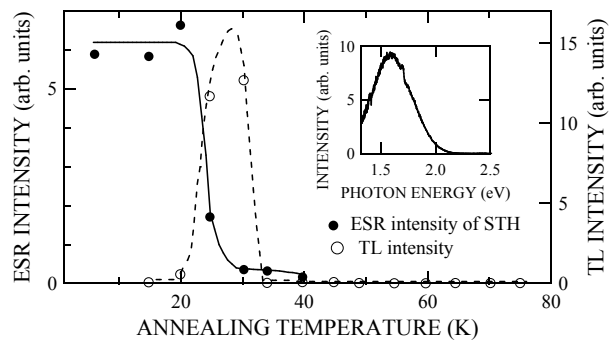}%
\caption{ESR intensity of the STH signals (closed circle) and 
total TL intensity during pulse annealing (open circle). 
The abscissa stands for pulse-annealing temperature. 
The ESR signals after pulse annealing were measured at 6 K. 
Solid and dashed lines are drawn for guides to the eye. 
The inset displays the TL spectrum under pulse annealing at 22 K.}
\label{STHvsTL}
\end{figure}

Figure \ref{STHvsTL} shows the thermal stability of the STH signals 
(closed circle) and the total TL intensity during the pulse annealing 
(open circle). 
The ESR signals after pulse annealing were measured at 6 K. 
The intensity of the STH signals sharply decreases above 20 K 
and the signals were hardly observed after pulse annealing at 40 K. 
The ESR signals in 280-340 mT also decrease up to 40 K by the similar amount 
to the STH signals, and keep almost constant under pulse annealing 
at 40-100 K. 
TL appears corresponding to the quenching of the STH signals over 20-30 K. 
The TL spectrum under pulse annealing at 22 K is displayed in the 
inset of Fig.\ \ref{STHvsTL}. 
The TL spectrum at 1.6 eV is spectrally in agreement to the red PL band; 
it is induced under excitation into the fundamental absorption 
region and increases in 20-30 K instead of the decrease of 
blue-green PL band at 2.55 eV.~\cite{Iwanaga1,Iwanaga2} 

\begin{figure}
\includegraphics[height=46.5mm,width=80mm]{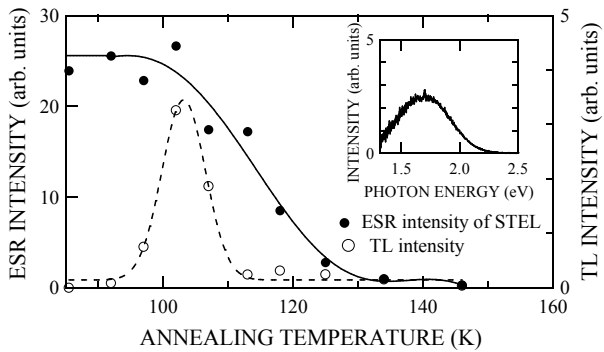}%
\caption{ESR intensity of the STEL signals (closed circle) and 
total TL intensity during pulse annealing (open circle). 
The abscissa stands for pulse-annealing temperature. 
The ESR signals after pulse annealing were measured at 5 K. 
Solid and dashed lines are drawn for guides to the eye. The inset displays 
the TL spectrum measured around the peak of the TL curve (dashed line).}
\label{STELvsTL}
\end{figure}

Figure \ref{STELvsTL} shows the stability of the STEL signals (closed circle) 
and the total TL intensity during the pulse annealing (open circle). 
The ESR signals after pulse annealing were measured at 5 K. 
The intensity of the STEL signals decreases above 100 K 
and the signals are quenched at 145 K. 
The ESR signals in 280-340 mT also vanish together with the STEL signals. 
TL was observed over 90-140 K, strongly in 100-105 K. The TL spectrum 
measured around the peak of the TL curve is displayed in the inset 
of Fig.\ \ref{STELvsTL}; the TL spectrum peaks at 1.7 eV and is located in 
the high-energy side in comparison with the spectrum in Fig.\ \ref{STHvsTL}. 
The TL spectrum in Fig.\ \ref{STELvsTL} 
is spectrally in agreement with the PL spectrum in this 
temperature range. The PL has the extrinsic nature;~\cite{Iwanaga4} 
it is induced under excitation even in the energy range 
($\,\hbar\omega\ge 3.5$ eV) lower than the fundamental absorption edge 
of 3.8 eV.~\cite{Iwanaga2}

\section{SPIN-HAMILTONIAN ANALYSIS\label{analysis}}
\subsection{Structure of self-trapped electrons\label{AnaSTEL}}
The ESR signals prominently appeared in 410-530 mT have the 
$g$ value of $g<2$ and therefore are ascribed to electron 
centers.~\cite{Slichter} 
They have the similar rotation-angle dependence with each other as shown in 
Fig.\ \ref{STEL}, and thermally disappear together. 
The intensity ratio seems to reflect the isotope effect. Indeed, 
Pb$^{2+}$ ions consist of nuclear spin $I=0$ 
(contained 79\% naturally) and $I=1/2$ (contained 21\% naturally). 
The pair of two-equivalent Pb-ions has the total nuclear spin of $0$ 
(singlet), $1/2$ (doublet), and $1$ (triplet); the constituent ratio is 
$19.2:10:1$. When the degeneracy of the multiplets is lifted, the ratio 
of the singlet to the two doublets is $19.2:5:5\approx 4:1:1$. 
This ratio is in agreement with the ratio $4:0.9:1$ estimated 
from Figs.\ \ref{ESR} and \ref{STEL}, 
and moreover it is probable that the far weak signals fitted with dashed 
lines in Fig.\ \ref{STEL} come from the nondegenerated triplets. 

Therefore, one can assume that an electron equivalently localizes on 
two nearest-neighbor Pb$^{2+}$ ions and they form a Pb$_2$$^{3+}$ center. 
The spin Hamiltonian $\mathcal{H}$ with Zeeman and hyperfine terms 
is given by 
\begin{equation}
\mathcal{H}=\mu_{B}{\bf S}\cdot \underline{g}\cdot {\bf B} + 
g_{0}\mu_{B}{\bf I}_1\cdot  \underline{A_1}\cdot {\bf S} + 
g_{0}\mu_{B}{\bf I}_2\cdot  \underline{A_2}\cdot {\bf S},\label{Ham}
\end{equation}
where $\mu_{B}$ denotes the Bohr magneton, 
{\bf S} the electron spin, $\underline{g}$ the Zeeman tensor, 
{\bf B} magnetic field vector, $g_{0}$ the free-electron $g$ factor, 
{\bf I}$_i$ the nuclear spin, and 
$\underline{A_i}$ the hyperfine tensor ($i=1, 2$). 
The Pb-ion dimer does not contain the nuclear quadrupole term because 
$I_i = 0$ or $1/2$. 

Each rotation-angle series of the ESR signals in Fig.\ \ref{STEL} takes 
maximum and minimum values when $\mathbf{B}\parallel a$ and 
$\mathbf{B}\parallel b$. The dependence enables to set the principal 
$g$ and $A$ axes such as $x=c$, $y=b$, and $z=a$. 

Under axial symmetry $(A_x = A_y)$ and the second-order perturbation taking 
Zeeman term as the unperturbed term, 
the Hamiltonian (\ref{Ham}) is transformed, as shown explicitly in 
Ref.\ \onlinecite{Schoemaker}, into the equation 
describing the allowed ESR transitions. The axial symmetry implies that 
the dimer-molecular axis is chosen to be the $z$ axis. 

The modified equation well describes the rotation-angle dependent series of 
the ESR signals around the $a, b$ and $c$ axes as in Fig.\ \ref{STEL}. 
The principal $g$ values are obtained by fitting the Zeeman series of 
$I = 0$; the solid line with index $(I, M_I)=(0, 0)$ in 
Fig.\ \ref{STEL} represents the fitted Zeeman line. 
Principal $A$ values are obtained by fitting the series of 
$I = 1/2$, using the principal $g$ values; the 
solid lines with indexes $(1/2, 1/2)$ and $(1/2, -1/2)$ in Fig.\ \ref{STEL} 
fit the series by varying the principal $A$ values only. The 
dashed lines in Fig.\ \ref{STEL} are drawn with the $g$ and the $A$ values, 
and those with $(1, 0)$ and $(1, -1)$ describe the series well. 
This analysis confirms that the ESR spectrum named as STEL in 
Fig.\ \ref{ESR} and the ESR signals in Fig.\ \ref{STEL} originate from the 
electron-trapping center of Pb$_2$$^{3+}$. 
As for the electron centers, it is improbable that they are affected by 
vacancies or impurities because the rotation-angle dependence 
excludes the symmetry breaking around the electron centers 
by the permanent lattice defects. 
The coexistence of STEL's and ``perturbed'' STEL's by the defects 
is also excluded because of the single rotation-angle pattern and 
the thermal decay. 
Therefore, the electron centers named as STEL in Fig.\ \ref{ESR} are 
declared to be self-trapping electron centers of Pb$_2$$^{3+}$. 

\begin{figure}
\includegraphics[height=61.1mm,width=55mm]{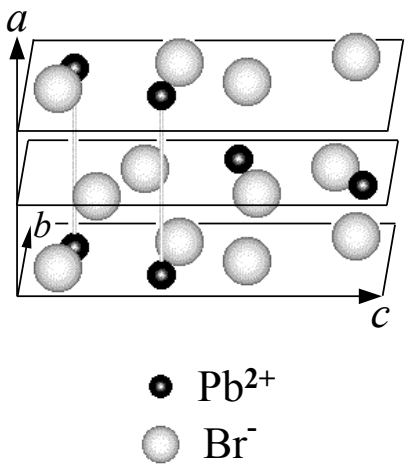}%
\caption{Schematic configurations of the electron-self-trapping 
Pb$_2$$^{3+}$ center in PbBr$_2$ crystal. 
Four Pb$^{2+}$ ions (small black ball) and eight Br$^-$ ions 
(large white ball) in the two nearest-neighbor $bc$ planes 
constitute the unit cell of the crystal. The lengths of the unit cell 
along the $a$, $b$, and $c$ axes are 4.767, 8.068, and 9.466 \AA, 
respectively (Ref.\ \onlinecite{Wyckoff}). 
There exist the two configurations of the STEL centers in the 
crystal, and each STEL center of Pb$_2$$^{3+}$ is schematically represented 
with the two nearest-neighbor Pb$^{2+}$ ions along the $a$ axis and 
the stick which depicts the covalent bond of the two Pb$^{2+}$ ions via 
the electron. The spin-Hamiltonian parameters of the Pb$_2$$^{3+}$ center 
are listed in Table \ref{STELparameters}.}
\label{STELconfig}
\end{figure}

The structures of the STEL centers are schematically shown in 
Fig.\ \ref{STELconfig}; among the two nearest-neighbor Pb-ion pairs, 
the pairs along the $a$ axis are the configurations consistent with 
the analysis unfolded in this section. In Fig.\ \ref{STELconfig}, each stick 
which bonds the two Pb$^{2+}$ ions stands for the covalent bond via 
the localized electron. As described in Sec.\ \ref{ESRresult}, the broad 
line width of the STEL signals probably hides the twofold splits in the 
$ab$ and the $ac$ planes observed in PbCl$_2$ (Refs.\ \onlinecite{Nistor1} 
and \onlinecite{Hirota}), and consequently the $x$ and $y$ axes 
determined here may deviate slightly from the principal $x$ and $y$ axes 
corresponding to each of the twofold splits. 

The principal values of Zeeman $g$ and hyperfine $A$ tensors are listed 
in Table~\ref{STELparameters}. The principal $g$ values vary from 
1.4-1.7 and are comparable to the values of Pb$_2$$^{3+}$ in PbCl$_2$ and 
NaCl:Pb.~\cite{Nistor1,Hirota,Heynd} 
This analysis only provides the absolute values of $A_x$, $A_y$ 
and $A_z$, and the principal $A$ values are in the same order of several 
tens mT as in PbCl$_2$ and NaCl:Pb. 
In accordance with Pb$_2$$^{3+}$ in PbCl$_2$ and NaCl:Pb, 
the same signs of principal $A$ values are chosen for 
Pb$_2$$^{3+}$ in PbBr$_2$. The parameters $\rho_{s}$ and $A_{\sigma}$ of 
PbBr$_{2}$ are calculated with the $g$ and the $A$ values in the 
perturbative analysis for the electronic levels of dimer-molecular 
centers.~\cite{Schoemaker2,Heynd} The physical meanings of the 
$\rho_{s}$ and the $A_{\sigma}$ are discussed in Sec.\ \ref{signs}. 

\begin{table*}
\caption{Spin-Hamiltonian parameters of Pb$_2$$^{3+}$ electron centers. 
Principal $A_i$ values ($i=x, y, z$), 
$\rho_s$, and $A_{\sigma}$ are represented in mT. 
The meanings of $\rho_s$ and $A_{\sigma}$ are discussed in 
Sec.\ \ref{signs}. 
The parameters of Pb$_2$$^{3+}$ centers in PbCl$_2$ and NaCl:Pb are also 
cited for comparison.}
\label{STELparameters}
\begin{ruledtabular}
\begin{tabular}{c|cccccccc}
Electron center & $g_x$ & $g_y$ & $g_z$ & 
$A_x$ & $A_y$ & $A_z$ & $\rho_s$ & $A_{\sigma}$ \\
\colrule
Pb$_2$$^{3+}$ & 1.563 & 1.416 & 1.690 & 
$-87$ & $-87$ & 100 & 39 & $-14$ \\
in PbBr$_2$\footnote{This work; Sec.\ \ref{AnaSTEL}.} 
& $\pm$0.002 & $\pm$0.002 & $\pm$0.002 & $\pm$4 & $\pm$3 & $\pm$5 & & \\
\colrule
Pb$_2$$^{3+}$ & 
1.549 & 1.379 & 1.718 & $-85$ & $-85$ & 109 & 39 & $-5$ \\
in PbCl$_2$\footnote{Reference \onlinecite{Nistor1}.} & 
$\pm$0.001 & $\pm$0.001 & $\pm$0.003 & $\pm$1 & $\pm$1 & $\pm$1 & & \\
\colrule
Pb$_2$$^{3+}$ & 
1.494 & 1.322 & 1.662 & NR\footnote{Not reported.} & NR & NR & NR & NR \\
in PbCl$_2$\footnote{Reference \onlinecite{Hirota}.} & 
$\pm$0.005 & $\pm$0.005 & $\pm$0.005 & & & & & \\
\colrule
Pb$_2$$^{3+}$ & 
1.469 & 1.300 & 1.621 & $-115$ & $-123$ & 115 & 46 & $-29$ \\
in NaCl:Pb\footnote{References \onlinecite{Nistor1} and \onlinecite{Heynd}.}& 
$\pm$0.002 & $\pm$0.002 & $\pm$0.002 & $\pm$1 & $\pm$1 & $\pm$1 & & \\
\end{tabular}
\end{ruledtabular}
\end{table*}

\subsection{Structures of self-trapped holes\label{AnaSTH}}
The three prominent ESR signals in 350-390 mT named as STH in Fig.\ \ref{ESR} 
belong to the series with the similar rotation-angle dependence and the same 
thermal stability under pulse annealing. The 
intensity of the ESR signals in 280-340 mT decrease up to 40 K 
by the similar amount to the three signals in 350-390 mT. 
Therefore, there probably exist further signals in 280-340 mT 
of the same origin with the three signals in 350-390 mT, though the 
overlap with other signals prevents from observing them separately. 
The ESR signals in 280-390 mT have the $g$ values of $g\approx 2$ and 
probably originate from hole centers.~\cite{Slichter} 

In PbBr$_2$, we assume two-equivalent-nuclei Br$_2$$^-$ 
as the hole center with more than three resonances in 280-390 mT. 
This assumption is consistent with the ratio of $3:2:1$ estimated in 
Fig.\ \ref{ESR}; full signals of Br$_2$$^-$ would be composed of seven 
resonances with the ratio of $1:2:3:4:3:2:1$ in the first-order hyperfine 
effect associated with the Br-nuclear spin of $3/2$. 

The spin Hamiltonian $\mathcal{H}$ is also expressed by eq.\ (\ref{Ham}). 
For Br$_2$$^-$, quadrupole term can be added as a second-order 
effect~\cite{Schoemaker} because the nuclear spin of Br$^-$ is $3/2$. 
However, explicit second-order effects have not been observed in our 
measurements. 
Therefore, we apply the eq.\ (\ref{Ham}) containing only first-order term 
to the experimental data and moreover hypothesize the agreement of principal 
$g$ and $A$ axes. 

As shown in Fig.\ \ref{STH}, the solid-line and the dashed-line series are 
mirror symmetric for the $bc$ plane, and the rotation-angle dependences 
imply that the principal $g$ and $A$ axes deeply tilt from any 
crystallographic axis. This deviation and the overlap of the signals in 
280-340 mT make it difficult to determine the orientation of principal 
$x$, $y$, and $z$ axes. 
However, it is possible only to determine approximately the orientation 
of $z$ axis from the well-resolved signals in 350-390 mT. 
Thus, the first and second Euler angles are 
respectively 51$^{\circ}$ and 126$^{\circ}$ for the series fitted with the 
solid lines in Fig.\ \ref{STH}, and those are 129$^{\circ}$ and 
126$^{\circ}$ for the dashed lines. 
The third angle cannot be determined experimentally. 
We assume the angle to be 0$^{\circ}$, and consequently the orientation of 
$x$ and $y$ axes determined here is not necessarily in agreement with 
the principal $x$ and $y$ axes of Br$_2$$^-$ centers. 

\begin{figure}
\includegraphics[height=66.7mm,width=55mm]{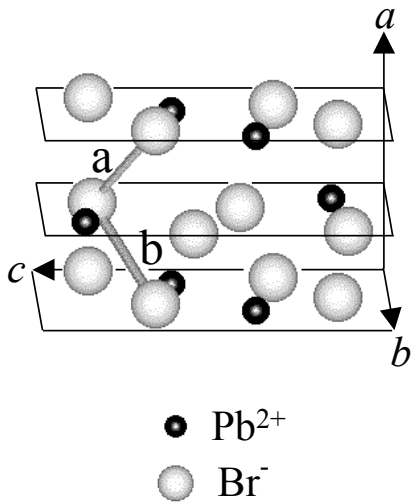}%
\caption{Schematic configurations of the hole-self-trapping Br$_2$$^-$ 
centers in PbBr$_2$ crystal. 
Br$_2$$^-$ is composed of two Br$^-$ ions (large white ball) and one hole. 
Each stick schematically represents the covalent bonds 
of two Br$^-$ ions via the hole. The two dimer-molecular Br$_2$$^-$ 
named as a and b respectively correspond to the solid-line and 
the dashed-line series in Fig.\ \ref{STH}. 
Two Pb$^{2+}$ ions (small black ball) and four Br$^-$ ions are sited 
in each $bc$ plane. The spin-Hamiltonian parameters of the Br$_2$$^-$ 
centers (the pairs a and b) are listed in Table \ref{STHparameters}.}
\label{STHconfig}
\end{figure}

The spin-Hamiltonian analysis is performed in a similar procedure described 
in Sec.\ \ref{AnaSTEL}, except for $I = 3$. 
As shown in Fig.\ \ref{STH}, the Hamiltonian well describes the 
rotation-angle dependence of the three prominent signals in 350-390 mT and 
some other prominent signals in the region lower than 340 mT. 
Table \ref{STHparameters} presents the principal $g$ and $A$ values 
obtained by the fitting and shows that the values of Br$_2$$^-$ 
centers in PbBr$_2$ are comparable to those in alkali bromides. 
Though this spin-Hamiltonian analysis cannot provide the signs of 
hyperfine parameters $A_x$, $A_y$, and $A_z$, 
the molecular analysis~\cite{Schoemaker2} of electronic levels 
in Br$_2$$^-$ centers requires the signs to be positive,  
and the same signs are chosen in PbBr$_2$ in accordance with 
the analysis. The parameters $\rho_{s}$ and $A_{\sigma}$ in Table 
\ref{STHparameters} are discussed in comparison with those of other host 
crystals in Sec.\ \ref{signs}. 

The Br-ion pairs whose axis is in agreement with the principal $z$ axis of 
the solid-line series in Fig.\ \ref{STH} are limited to the pair a 
in Fig.\ \ref{STHconfig}; for the pair a, the first and second Euler 
angles of the molecular axis are respectively 50.7$^{\circ}$ and 
126.1$^{\circ}$ when the lattice ions are located at the equilibrium 
positions. Similarly, the dashed-line series in Fig.\ \ref{STH} correspond 
to the pair b in Fig.\ \ref{STHconfig}; the first and second Euler angles 
are 129.3$^{\circ}$ and 126.1$^{\circ}$, respectively. 
The pairs a and b are indeed mirror symmetric for the $bc$ plane. 

The good agreement of the dimer-molecular axes 
with the direction of the Br-ion pairs in the crystal confirms 
that the STH signals originate from the Br$_2$$^-$ centers which are not 
affected by any permanent lattice defect. 
Therefore, the hole centers named as STH in Sec.\ \ref{ESRresult} 
are ascribed to the self-trapping hole centers of Br$_2$$^-$. 

\begin{table*}
\caption{Spin-Hamiltonian parameters of Br$_2$$^-$ hole centers. 
Principal $A_i$ ($i = x, y, z$) values, $\rho_s$, and $A_{\sigma}$ are 
represented in mT. The meanings of $\rho_s$ and $A_{\sigma}$ are discussed 
in Sec.\ \ref{signs}. The parameters of Br$_2$$^-$ centers 
in KBr, NaBr, and RbBr are cited for comparison.}
\label{STHparameters}
\begin{ruledtabular}
\begin{tabular}{c|cccccccc}
Hole center & $g_x$ & $g_y$ & $g_z$ & $A_x$ & $A_y$ & $A_z$ & 
$\rho_s$ & $A_{\sigma}$ \\
\colrule
Br$_2$$^-$ (pair a)\footnotemark[1] 
& 2.16 & 2.17 & 1.90 & 10 & 10 & 46 & 16 & 17 \\
in PbBr$_2$ 
& $\pm$0.02 & $\pm$0.02 & $\pm$0.01 & $\pm$4 & $\pm$4 & $\pm$1 & & \\
\colrule
Br$_2$$^-$ (pair b)\footnotemark[1] 
& 2.18 & 2.23 & 1.90 & 10 & 10 & 46 & 19 & 14 \\
in PbBr$_2$ 
& $\pm$0.05 & $\pm$0.05 & $\pm$0.03 & $\pm$1 & $\pm$1 & $\pm$7 & & \\
\colrule
Br$_2$$^-$ in NaBr\footnotemark[2] 
& 2.1514 & 2.1968 & 1.9791 & 6.50 & 7.38 & 43.1 & 18.23 & 12.66 \\
Br$_2$$^-$ in KBr\footnotemark[2] 
& 2.1629 & 2.1623 & 1.9839 & 7.68 & 7.66 & 45.0 & 17.98 & 14.45 \\
Br$_2$$^-$ in RbBr\footnotemark[2] 
& 2.1683 & 2.1524 & 1.9846 & 8.36 & 8.12 & 45.5 & 17.80 & 15.13 \\
\end{tabular}
\end{ruledtabular}
\footnotetext[1]{This work; Sec.\ \ref{AnaSTH}.}
\footnotetext[2]{Reference \onlinecite{Schoemaker2}; the 
accuracies of parameters $g$ and $A$ are within $\pm 0.0005$ and 
$\pm 0.07$, respectively.}
\end{table*}

\section{DISCUSSION\label{discussion}}
\subsection{Correlation of self-trapped states with luminescence%
\label{STSvsLumi}}
Thermal stability of the ESR signals indicates that the STEL's and 
the STH's coexist only in the temperature range below 30 K. 
Therefore, intrinsic luminescence related to both STEL's and STH's 
would appear below 30 K. 

Blue-green (BG) PL band at 2.55 eV is the dominant intrinsic PL 
below 20 K in PbBr$_2$, and the phosphorescent decay is well described by 
the radiative recombination model.~\cite{Iwanaga2} 
The model assumes the process that the spatially separated $e$-$h$ pair 
gets close by tunneling motion, forms a self-trapped exciton (STE), 
and recombines with radiation. 
The phosphorescent decay suggests that the BG-PL band originates from 
the distant pairs of a STEL and a STH. 

The BG-PL band thermally decreases in 20-30 K and disappears above 
30 K.~\cite{Iwanaga2} 
The thermally stable range of the BG-PL band corresponds to  
that of both STEL's and STH's. 
These correspondence supports that the BG-PL band is closely related to 
both STEL's and STH's. 
In particular, the BG-PL band is not induced in the temperature range above 
30 K, where the STH's disappear as shown in Fig.\ \ref{STHvsTL}. 
Therefore, it is plausible that the STE's yielding the 
BG-PL have the configuration of (Br$_2$$^-$ + electron), though 
it is unknown whether the STE is the nearest-neighbor pair 
of a STH and a STEL or has the configuration that a Br$_2$$^-$ center bounds 
an electron in the excited orbital. 
On the other hand, the red PL band grows instead of the quenching of the 
BG-PL band in 20-30 K; as shown in Fig.\ \ref{STHvsTL}, 
the temperature range corresponds to the growth 
of the red TL that has the same shape with the red PL band. 
Because of the instability of the STH centers in 20-30 K, the red PL 
and the red TL are ascribed to the STE's associated with 
the STEL centers of Pb$_2$$^{3+}$ stable in 20-30 K. 
To clarify the configurations of the self-trapped excitons directly, 
the examination with optically detected magnetic resonance technique 
is preferable, because this ESR study can only suggest the configurations 
of the STE's yielding the BG-PL and the red PL bands. 

As for the thermal stability of self-trapping centers, it is to be noted 
that the temperature where the STH's and the STEL's disappear under pulse 
annealing does not necessarily indicate that of the thermal quenching of the 
self-trapping centers itself. Indeed, the thermal decay of Br$_2$$^-$ 
in alkali halides is affected by the thermal activation of trapped 
electrons and positive vacancies.~\cite{Schoemaker2}
In PbBr$_2$, vacancies of Br$^-$ and Pb$^{2+}$ inevitably exist 
because of the high ionic conductivity,~\cite{Verwey2,Ober} 
and the density of Br$^-$ vacancy is estimated at 
more than 10$^{17}$ cm$^{-3}$. 
Consequently, it is likely that 
the thermal activation of vacancies affects the STH's: 
(i) it makes the self-trapped states unstable, induces the thermal transfer 
of the holes, and leads the holes to radiative and/or nonradiative decay, or 
(ii) it deforms the self-trapping centers to other trapping centers 
associated with vacancies. 
The thermal stability of the STEL's may be also affected by the thermal 
activation of other hole centers as discussed in Sec.\ \ref{Other}. 

\subsection{Electronic structures of self-trapped centers\label{signs}}
In this section we tentatively apply the perturbative analysis 
in Refs.\ \onlinecite{Schoemaker2} and \onlinecite{Heynd} 
to the dimer centers, and 
discuss the electronic structures of STEL's and STH's in PbBr$_2$ 
from the comparison with other crystals. 

According to the analysis,~\cite{Heynd,Schoemaker2} 
the mixing by spin-orbit interaction 
between the ground $\sigma_g$ and the excited orbitals $\pi_g^x$, $\pi_g^y$ 
in the dimer molecule of Pb$_2$$^{3+}$ and Br$_2$$^-$ 
are calculated from the $g$ shift and the principal $A$ values 
under the Hartree-Fock approximation and the second-order perturbation 
taking the spin-orbit interaction as the perturbed term. The 
parameters $\rho_s$ in Tables \ref{STELparameters} and \ref{STHparameters} 
denote the dipole-dipole interaction between the magnetic moments of the 
electron (or hole) and nuclei, and represent the anisotropic contribution 
to the hyperfine term. 
The parameters $A_\sigma$ represent the isotropic contribution to the $A$ 
tensor and are given by 
$A_{\sigma}=A_{\sigma}^{s}+A_{\sigma}^{e}$ where $A_{\sigma}^{s}$ 
is the positive Fermi contact term and $A_{\sigma}^{e}$ the negative 
exchange-polarization term of the inner electrons. 

For Pb$_2$$^{3+}$, $A_{\sigma}^{s}$ is 
proportional to $6s$-state mixing into the ground orbital 
$\sigma_g = \alpha_g (6s_1 + 6s_2) + \beta_g (6p_{1,z} - 6p_{2,z})$ 
where $\alpha_g$ and $\beta_g$ are constants, and $6s$ and $6p$ respectively 
denote the atomic wave functions of the $6s$- and $6p$-states in 
Pb$^{2+}$. $\rho_s$ is $\rho_s \propto \beta_g{}^2 \langle r^{-3}\rangle$ 
where $\langle r^{-3}\rangle$ is the 
spatial distribution of the electron, and the small 
$\rho_s$ values result in the spread of the electron as long as the 
$\beta_g$'s of Pb$_2$$^{3+}$ are comparable in different host crystals. 
As shown in Table \ref{STELparameters},  
$\rho_s$ in PbBr$_2$ has the same value with that in PbCl$_2$ and 
15\% smaller than that in NaCl:Pb; 
the $\rho_s$ suggests that the distribution of STEL in PbBr$_2$ 
spreads as outward as in PbCl$_2$ and more outward than in NaCl:Pb. 
$A_{\sigma}$ in PbBr$_2$ is negatively 2.8 times larger than that 
in PbCl$_2$. In other words, the positive part $A_{\sigma}^{s}$, which 
reflects the $6s$-mixing into the ground orbital $\sigma_g$, contributes 
more than the negative $A_{\sigma}^{e}$ in PbCl$_2$. 
This result qualitatively suggests that the $6s$-mixing into 
the $\sigma_g$ is larger in PbCl$_2$ than in PbBr$_2$. 
The difference is probably ascribed to the components of 
the top of valence band. The valence-band structure of PbCl$_2$ 
with more mixing of the $6s$-states in Pb$^{2+}$ ions may make it possible 
for the holes to form the STH centers of Pb$^{3+}$ as observed in PbF$_2$ 
(Ref.\ \onlinecite{Pb3+2}). Indeed, the mixing ratios of the $6s$-states 
into the top of the valence bands become large in order of 
PbBr$_2$, PbCl$_2$, PbF$_2$ (Ref.\ \onlinecite{Fujita}). 

The complementary discussion can be applied to Br$_2$$^-$. 
Parameters $\rho_s$ and $A_{\sigma}$ in Table \ref{STHparameters} 
are comparable to each other. This analysis implies that 
the distribution of STH and the orbital functions of 
the $\sigma_g$ in PbBr$_2$ are quite similar to those in alkali bromides. 

\subsection{Other trapped centers associated with lattice defects%
\label{Other}}
The ESR signals in 280-340 mT have the $g$ values of $g\ge 2$, 
and the overlap of the signals prevents from 
discriminating each signal and analyzing with spin Hamiltonian. 
However, they vanish together with the STEL signals around 120 K 
under pulse annealing, and then the crystal emits TL as shown in 
Fig.\ \ref{STELvsTL}. 
Therefore, the signals in 280-340 mT are most likely to originate from 
hole-trapping centers. 

From the valence-band structure, Pb$^{3+}$ center can be considered 
as another candidate for intrinsic hole-trapping center, 
but the satellites, which stem from the isotope ${}^{207}$Pb of $I=1/2$ 
(contained 21\% naturally), have not been observed in 0-1 T 
just as seen in PbF$_2$ crystals~\cite{Pb3+2} and Pb-doped KCl 
crystals.~\cite{Pb3+1} 
Interstitial lattice defects, namely, Frenkel defects have been also excluded 
in PbBr$_2$.~\cite{Verwey2,Ober}
Consequently, the signals in 280-340 mT except for the unresolved 
STH signals probably originate from the extrinsic hole center 
associated with the permanent lattice defects such as vacancy or 
impurity or both. The extrinsic hole-trapping centers are efficient 
competitors of the STH centers because they survive above 100 K 
together with not a few STEL and 
disappear around 120 K with emitting TL. 

\section{CONCLUSIONS}
Self-trapping electron and hole centers are simultaneously 
photoinduced by two-photon interband excitation with 120-fs-width laser 
light at 3.10 eV and have been directly detected with ESR technique. 
The excitation enables to induce $e$-$h$ pairs with enough density for 
the present ESR measurements. 
Spin-Hamiltonian analyses have revealed that the structures of STEL's and 
STH's are the dimer-molecular Pb$_2$$^{3+}$ and Br$_2$$^-$, 
respectively. The STEL centers of Pb$_2$$^{3+}$ orient 
along the crystallographic $a$ axis. The holes select 
self-trapping sites among many nearest-neighbor Br-ion pairs in the unit 
cell, and form the STH centers of Br$_2$$^-$ with the two particular 
orientations. 
From the comparison of the principal $g$ and $A$ values, the Pb$_2$$^{3+}$ 
and the Br$_2$$^-$ centers are similar to those in other crystals 
studied so far. 

From this ESR study, we conclude that 
both electrons and holes in PbBr$_{2}$ crystals  strongly interact with 
acoustic phonons and do relax into the individual self-trapped states. 

Moreover, the STEL's and the STH's centers coexist below 30 K, and 
the STH centers disappear in 20-30 K. 
The temperature range is in good agreement with the range where 
the phosphorescent blue-green PL band at 2.55 eV is induced. 
The agreement supports the conclusion of the previous PL study 
(Ref.\ \onlinecite{Iwanaga2}) that both STEL and STH centers are 
related to the radiative recombination process yielding 
the blue-green PL band. 

\bibliography{iwanaga}

\end{document}